\documentclass[
aps,nofootinbib,
showpacs,showkeys,preprint
tightenlines,preprintnumbers,] {revtex4}

\usepackage{epsf,epsfig,subfigure,
graphicx,amsmath,amssymb}
\usepackage{color}

\newcommand{\dis}[1]{\begin{equation}\begin{split}#1\end{split}\end{equation}}

\newcommand{\gev}{\,\textrm{GeV}}
\newcommand{\mev}{\,\textrm{MeV}}

\newcommand{\eV}{\,\textrm{eV}}

\newcommand{\LQCD}{\Lambda_{\rm QCD}}

\def\eVV{\,\textrm{eV}}

\def\etal{{\it et al}}
\def\ie{{\it i.e.~}}

\def\thb{\bar{\theta\,}}

\def\Mp{M_{\rm P}}

\def\EE8{E$_8\times$E$_8^\prime$}

   \def\qg{q\&g}

\begin{document}

\title{Axion energy density,
bottle neck period, and $\thb$ ratios between early and late times} 
   
\author{Jihn E.  Kim,$^{1,2}$ Se-Jin  Kim,$^2$ Soonkeon Nam$^2$ }
\address
{ 
$^{1}$Center for Axion and Precision Physics Research (Institute of Basic Science), KAIST Munji Campus, Munjiro 193,  Daejeon 34051, Republic of Korea,  \\ 
$^{2}$Department of Physics, Kyung Hee University, Seoul 02447, Republic of Korea 
} 
 
\begin{abstract} 
The possibility of the ``invisible'' axion being cold dark matter relies on the acceptable estimates of the current axion energy density. The estimate depends on the nature of QCD phase transition at a few hundred MeV and the evolution of the misalignment angle $\thb$. The onset of $\thb$ oscillation undergoes a bottleneck period which occurred during the QCD phase transition. In addition, the anharmonic coupling of order $a^4$ affects the $\thb$ evolution. From the time that the anharmonic effect is negligible, it is rather simple to calculate the ratio of  $\thb$'s between early and late times. For multi GHz oscillations, the current age of the Universe needs at least $10^{27}$ oscillations which limits an exact calculation of $\thb$. We establish a stepwise approximation for numerical solutions of the differential equation and obtain $\thb_{\rm now}/\thb_f\approx 3\times 10^{-17}$ for $m_a\simeq 10^{-4}\,\eVV$, where $t_f$ is the first time that the full hadronic phase (after the QCD phase transition) was established.

\end{abstract}
\maketitle


\section{Introduction}\label{sec:Introduction}

There have been increasing efforts to find out the QCD axion in the mass range  $10^{-5} \eVV\sim 10^{-4}\eVV$ \cite{JeongJ17,Jeong17,RedondoVenice}. This range corresponds to the so-callecd  ``invisible'' axion \cite{KSVZ1,KSVZ2,DFSZ1,DFSZ2} which has survived until now.  Usually, the axion search limit is presented in the axion-to-photon conversion rate ($\sigma_{a \to \gamma}\propto |c_{a\gamma\gamma}|^2$) versus the  axion mass $m_a$ (or the decay constant $f_a$) plane where the axion mass is the key parameter in determining the classical motion of the axion field. In these  plots, the theory curves are simply the relations between the axion-photon-photon coupling $|c_{a\gamma\gamma}|$ versus the axion mass and hence do not depend on the axion evolution history in the Universe. But, the experimental search limits in these plots are presented under the assumption that the axion contribution to the energy density of cold dark matter (CDM) in the Universe is about 27\% \cite{KimRMP10}. So, it is of utmost importance to know  the current axion energy density in the Universe as accurately as possible.

The shift symmetry direction $\thb$ (or $a/f_a$) is the direction of the global phase symmetry.  If it were an exact symmetry, the so-called Peccei-Quinn (PQ) symmetry \cite{PQ77}, then the corresponding Goldstone boson would be exactly massless but as is well-known all  global symmetries are always broken \cite{KimFate18}. For the PQ symmetry, it is broken only by gauge anomalies always including the QCD anomaly and not anywhere else, especially not including breaking terms in the  potential $V$. The breaking terms by the QCD anomaly gives mass to the axion as noted by Weinberg and Wilczek \cite{Weinberg78,Wilczek78}. The PQ shift symmetry is a chiral symmetry involving quark fields, $u_{L,R}\to e^{\pm i\alpha_u}u_{L,R}, d_{L,R}\to e^{\pm i\alpha_d}d_{L,R}$ (or simply $u\to e^{ i\alpha_u\gamma_5}u, d\to e^{  i\alpha_d\gamma_5}d$), etc. This chiral symmetry is respected by the kinetic energy terms of the quark fields but is broken by the quark mass terms and is proportional to $m_u\LQCD^3$.  Translated into the axion mass, the axion mass square is proportional to $m_u^2\LQCD^2$, which is an expression in the quark and gluon phase ($\qg$-phase). So, the axion mass  has been scrutinized  for a long time.  If one considers the hadronic phase ($h$-phase), the axion mass expression is proportional to the hadronic parameters $m_{\pi^0}$ and $f_{\pi^0}$ in which case also the mass should vanish in the chiral limit, \ie  in the limit $m_u\to 0$. Since any quark mass can be used for the chiral limit,  it vanishes as $m_um_dm_s/(m_um_d+m_um_s+m_dm_s)$  \cite{BardeenTye78,Baluni78}. Here, the light quark masses are the values above the QCD phase transition. The axion mass at zero temperature, \ie in the $h$-phase, must satisfy the chiral property without directly using the absolute values of the current quark masses but only their ratios   \cite{BardeenTye78,Baluni78}. So, the axion mass operators in the $\qg$-phase and $h$-phase are given by, in case of two quark flavors,\footnote{Consideration of the strange quark $s$ will change the result only at a 5\% level, and for simplicity we neglect $s$ in this paper. Equation (\ref{eq:AxPhmasses}) is for $\thb\simeq 0$, and the exact symmetry relation is given in \cite{KimKim18,KimPRP87}.}
\dis{
&\textrm{Quark and gluon phase with~}\LQCD:~\left(\frac{m_u}{1+Z}\right)^2\LQCD^2\,\left(\frac12 \thb^2\right),\\[0.5em]
&\textrm{Hadronic phase with~}f_{\pi^0}^2 m_{\pi^0}^2:~ \frac{Z}{(1+Z)^2}f_{\pi^0}^2 m_{\pi^0}^2\,\left(\frac12 \thb^2\right) ,\label{eq:AxPhmasses}
}
where $\thb=a/f_a$ and $Z=m_u/m_d$. Note that in the $\qg$-phase the single particle parameters such as the current quark masses and $\LQCD$ are used and in the $h$-phase parameters of the many-body condensed phenomena are used. Note that $Z$ is just a ratio descending from the high energy scale.
So, the evolution of the ``invisible'' axion energy density requires the knowledge on the QCD phase transition below 1 GeV until the completion of the phase transition from the $\qg$-phase to the $h$-phase.  Two of us considered this region carefully and concluded that the phase transition was completed by the time $t_f \approx  63\,\mu{\rm s}$ corresponding to the cosmic temperature $T_f\simeq 126\,\mev$ \cite{KimKim18}.

In axion cosmology, the onset temperature $T_1$ of axion field oscillation is the crucial parameter, which is determined by the condition $m_a(T_1)=3H(T_1)$ which is satisfied in the $\qg$-phase. Usually, $T_1$ is determined at around 1\,GeV, and the subsequent axion oscillation has been studied. In Ref. \cite{KimKim18}, it was argued that $T_1$ can be significantly different from 1\,GeV.

If $T_1$ is given, one can calculate $\thb$ evolution in the Universe.
There is the bottleneck period from $t_1$ (at cosmic temperature   $T_1$) to $t_{\rm osc}$ (at cosmic temperature   $T_{\rm osc}$) where    $t_{\rm osc}$ is some time when the anharmonic effect is negligible. The bottleneck period has been known before \cite{Lyth98,Bae08}.
$T_{\rm osc}$ is above the critical temperature $T_c\simeq 165\,\mev$
 below $1\,\gev$. From  $t_c$ (the time at the critical temperature   $T_c$), the phase transition from the $\qg$-phase to the $h$-phase begins, which is completed at  $t_f$ (at  temperature   $T_f$).

 In this paper, we study the axion evolution equation in the bottleneck period 
\dis{
 \ddot\theta+3H\dot\theta+m_a^2\sin\theta=0,\label{eq:diffeq}
}   
where $\thb=a/f_a$, and subsequent decrease factor of $\thb$.
In Sec. \ref{sec:ChSymmBr}, we set up the phase transition for chiral symmetry breaking and the decrease factor of $\thb$ for the later sections.  In Sec. \ref{sec:AxCDMbottle},  we introduce the bottleneck period analytically as much as possible, and then  present a numerical calculation of the $\thb$ evolution in the anharmonic regime until $t_f$.  In Sec. \ref{sec:tbar}, we present the $\thb$ evolution in the $h$-phase, from $t_f$ until now. Section \ref{sec:Conclusion} is a conclusion

\section{The chiral symmetry breaking}\label{sec:ChSymmBr}

It was considered that  the QCD  phase transition is of  the first order  in studies of axion cosmology \cite{DeGrand84,Bae08}, starting at cosmic time $t_i\simeq t_c$ .  In the lattice community, it seems that it was so. Long time ago, Ref.  \cite{Kogut88} concluded that a lattice calculation with two light quarks with mass $7.5\mev$ allows the first order phase transition.  About 10 years ago, Ref. \cite{Forcrand07} showed the first order phase transition in the region of two light quarks and massive strange quark in the $m_{u,d}-m_s$ phase diagram, where   it is stated explicitly, ``It is numerically well-established the phase transition  is the first order in the quenched limit, and there is strong numerinal evidence for first order in the chiral limit \cite{Kogut88}".
Within this first order phase transition idea, the following energy density and pressure of three pions in the $h$-phase were used in the MIT bag model  \cite{DeGrand84},
 \dis{
\rho_\pi=\frac{3\pi^2}{30} T^4,~P_\pi=B+\frac{3\pi^2}{90} T^4,
}
 where $B$ is the bag parameter.
 
 Recently, this view of the first order phase transition has been changed to cross over transition \cite{LosAlamos14} where it was observed that the susceptibility does not increase as the volume increase, and the critical temperature is cited as  $154\pm1\pm8\,\mev$.  Since the hint of the cross over transition appears around 164\,MeV  \cite{LosAlamos14}, we use the critical temperature  $T_c$ given in \cite{ICTP16,Sharma16,Boyarski16} 
\dis{
T_c=165\,\mev.\label{eq:TcLatt}
}

\subsection{Bubble formation rate}\label{subsec:Bformationrate}
\begin{figure}[!t]
\includegraphics[width=0.39\textwidth]{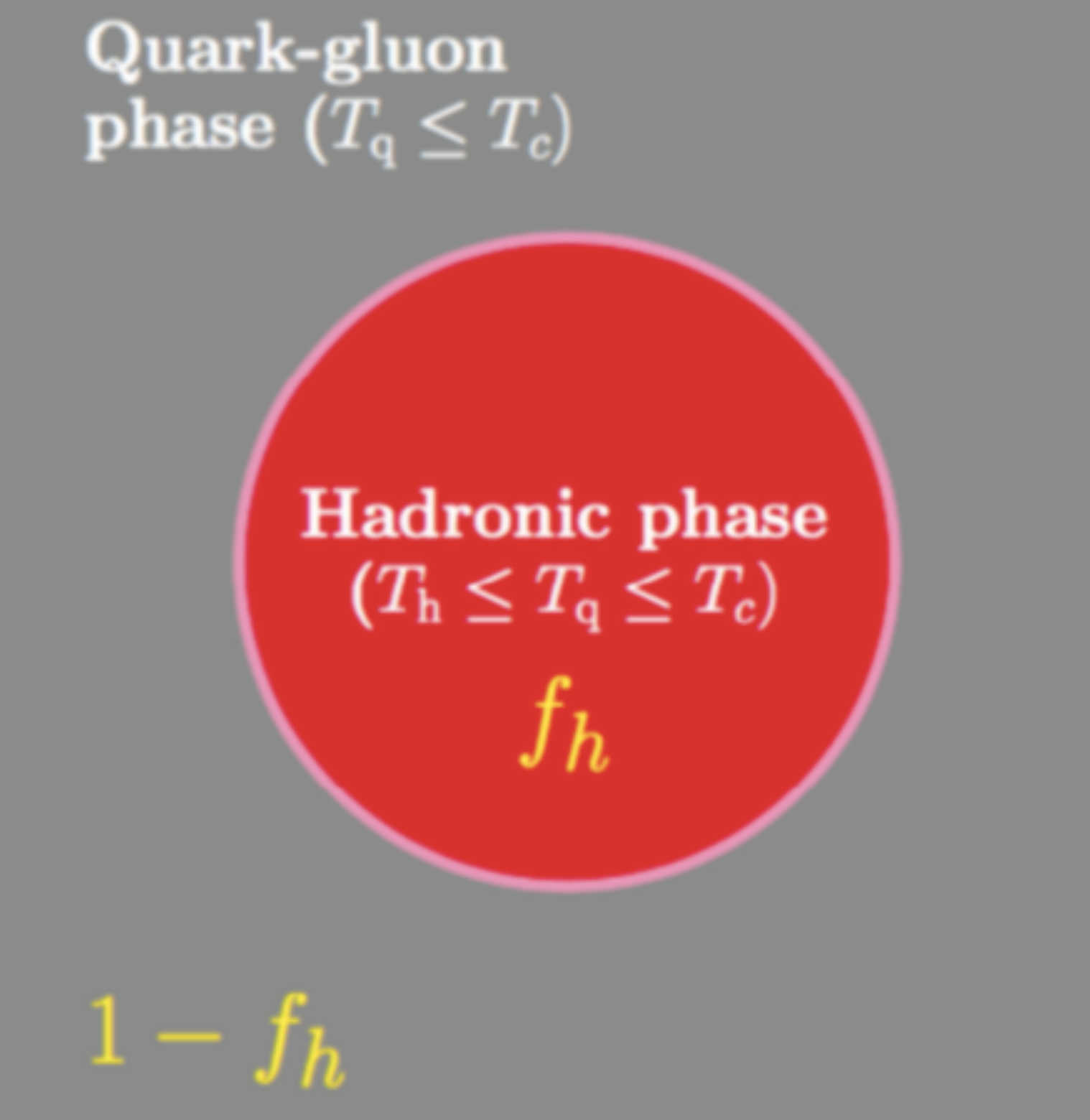} 
\includegraphics[width=0.4\linewidth]{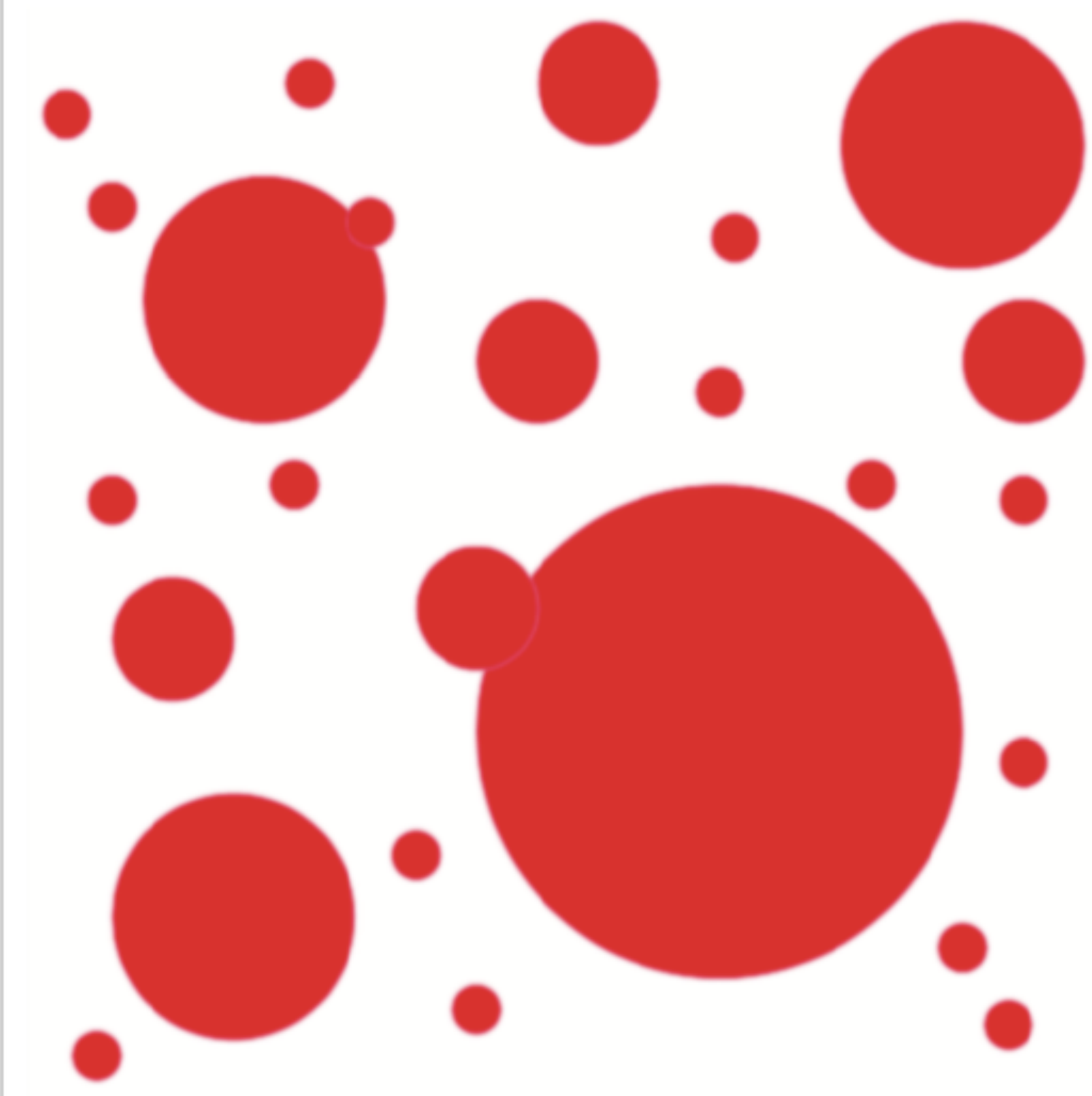} 
  \caption{Formation of hadronic bubbles at $T<T_c$. $f_h$ is the volume fraction of $h$-phase bubbles.  In the right figure, four scales of bubbles formed at four different time scales are shown.} \label{fig:hBubble} 
\end{figure}

In a humongous Universe, phase transitions are not likely to occur instantaneously but are assumed to be processed by formation of bubbles  and their expansion \cite{Okun74}. However, the field theoretic prescription \cite{Coleman77,CallCol77} is difficult to directly apply in the evolving Universe. Thus, this  phase transition was replaced by the tunneling idea with a phenomenological Lagrangian with parameter $\epsilon$ in \cite{KolbTurnerBk}.  Recently, in Ref. \cite{KimKim18} the bubble formation and expansion have been presented from the first principles, using the Gibbs free energy which is conserved during the  phase transition \cite{HuangStat}. The differential equation for the evolution of hadronic fraction $f_h$, starting from size $R_i$ was given as
\dis{
\frac{df_h}{dt}=\alpha(t) (1-f_h)+\frac{3}{ {(1+C f_h(1-f_h) )}(t+R_i)}f_h, \label{eq:diffequa}
}
where the initial condition is $f_h(t=0)=0$. Since one size of bubbles are created in this effective description, it is consistent with the cross over phase transition: in the beginning the second order phase transition creating a typical size $R_i$ is effective and later many size bubbles are present  as shown in right side of Fig. \ref{fig:hBubble}  behaving like the first order phase transition. 
The variables of Gibbs free energy are temperature and pressure. Conservation of Gibbs free energy in the phase transition requires the same temperature and pressure in the $\qg$- and $h$-phases; thus $h$-phase bubbles are formed at rest in the $\qg$-phase as shown in left side of Fig. \ref{fig:hBubble}. 
Conservation of the Gibbs free energy enables one to calculate the pion pressure at the critical temperature \cite{KimKim18}.  
 Knowledge of the pion pressure at $T_c$ is used to define a phenomenological differential equation of $f_h$ given in Eq. (\ref{eq:diffequa}), and  $\alpha(t)$ is calculated \cite{KimKim18}. The parameter $C$ effectively parametrizes the overlapping portions in the expansion of bubbles. The above differential equation determines the completion of the phase transition by the cosmic time $t_f\simeq   63\,\mu{\rm s}$ (at temperture $T_f\simeq 126\,\mev$).
 
We present $\thb$ evolution with the QCD phase transition idea discussed above. In the expanding RD Universe, the initial bubble radius $R_i$ is  increased to $R_i+t$, and the voume to  $ (R_i+t)^3$, and  the volume of the $h$-phase bubble expands with the rate $3/R(t)$ where $R(t)=t+R_i$. Therefore, including the Hubble expansion, the fraction $f_h$ of $h$-phase is approximately described by Eq. (\ref{eq:diffequa}).
The temperature of pions is equilibrated with collisions of leptons and photons and hence after the phase transition, we use the cosmic temperature $T$ rather than the pion temperature,  for the evolution in the $h$-phase. 
 
The time--temperature relation, in  seconds and MeV units  in the RD epoch, is approximated by
\dis{
T \simeq  \frac{5.07^{1/2} 10^{5}}{t_{\rm s}^{1/2}\mev^{1/2}}\mev= \frac{1}{t_{\rm MeV^{-1}}^{1/2}\mev^{1/2}}\mev,\label{eq:tTrelation}
}
such that at the radiation--matter equality point around $3.15\times 10^{11}$ s, the temperature is
\dis{
T_{\rm eq}\simeq  \frac{5.07^{1/2} 10^{5}}{t^{1/2}}\simeq  4.22~\eV, 
}
where the radiation energy is
\dis{
\rho_{\rm rad}\Big|_{\rm eq}=\rho_{\rm matter}\Big|_{\rm eq}\simeq g_*\frac{\pi^2}{30} T_{\rm eq}^4=3.54T_{\rm eq}^4\simeq 1.12\times 10^{-21}\mev^4,
}
where $g_*=43/4$ is used. Note that $H$ is given by
\dis{
H=\left(\frac{g_*}{3}\right)^{1/2}\frac{T^2}{\Mp}
\simeq 1.89\frac{T^2}{\Mp}.
}
Below $T_{\rm eq}$, the evolution of radiation and CDM have different powers on cosmic time $t$, but the $\thb$ evolution behaves always like CDM and we can calculate its development directly from $t_f$ to $t_{\rm now}$.

After $N$ axion oscillations from the time $t_f$, time increases to $t_f+Nm_a^{-1}$ and the temperature drops in the RD Universe by the factor
\dis{
\frac{ t_f^{1/2}}{(t_f+Nm_a^{-1})^{1/2}},
}
and the Hubble parameter becomes
 \dis{
&H_N(t) \simeq 1.89\frac{T_f^2 t_f^{1/2}}{\Mp(t_f+Nm_a^{-1}+t)^{1/2}},\\
&\dot{H}_N(t) \simeq - \frac{H_N(t)}{2(t_f+Nm_a^{-1}+t)}.
}

Let us  find out how much $\thb$ decreases in one oscillation.
In the linear regime, the evolution of $\thb$ satisfies
 \dis{
 L\thb\equiv \frac{d^2}{dt^2}\thb + m_a^2  \thb=- 3H \frac{d}{dt}\thb,~\thb=\theta_0+\thb_{\rm in}
 }
 where $L=(d/dt)^2+m_a^2$ and  $\theta_0$ is the solution without the $H$ term. We neglect the part of the solution of O($H^2$). The Green function $G(t-t')$ of $L$ satisfies $LG=\delta(t-t')$,  
 \dis{
 G(t-t')
&=\left\{ \begin{array}{l}
c_1 \cos m_a(t'-t) +c_2 \sin m_a(t'-t) ,~{\rm for~}t'<t\\
c_3 \cos m_a(t'-t) +c_4 \sin m_a(t'-t) ,~{\rm for~}t'>t\\
\end{array}, 
\right.\\
\frac{d}{dt'}G(t-t')
&=\left\{ \begin{array}{l}
-c_1 m_a\sin m_a(t'-t) +c_2 m_a\cos m_a(t'-t) ,~{\rm for~}t'<t\\
-c_3 m_a\sin m_a(t'-t) +c_4 m_a\cos m_a(t'-t) ,~{\rm for~}t'>t\\
\end{array},\right. ~{\rm with~the~condition}~  G'\Big|_{t-\epsilon}^{t+\epsilon}=1.
}
\dis{
c_4-c_2=m_a^{-1}.
}
The homegeneous solution $\theta_0$ starts from a cosine function at $t=0$ and we consider one oscillation, $\delta=\pi m_a^{-1}$. The boundary condition we choose is $\theta_0(0)=A$, \ie   $\theta_0(t)=A \cos m_at$ and $\dot\theta_0(t)=-m_aA\sin  m_at$.  The inhomogenious solution   $\thb_{\rm in}$ is given by  
\dis{
\thb_{\rm in}(t)  =&\int_0^\delta dt' \, (-3 H_N(t'))G(t-t')\dot\theta_0(t')\\
=&  \Big[\int_{0}^{t}dt'\, (-3H_N(t')) [c_1 \cos m_a(t'-t) +c_2 \sin m_a(t'-t)] (-  m_aA \sin m_at') \\
&   +\int_{t}^{ \delta}dt'\,  (-3H_N(t'))[c_3\cos m_a(t'-t) +(m_a^{-1}+c_2) \sin m_a(t'-t)](-m_aA \sin  m_at'  )\Big].
} 
 For a large $N\gg 1$,  $H_{N}(t')$ can be taken as a  constant within the short interval,
 \dis{
\thb_{\rm in}(t) =&3m_aAH_N(0)  \Big[\int_{0}^{t}dt'\, [c_1 \cos m_a(t'-t) +c_2 \sin m_a(t'-t)]\sin m_at' \\
&   +\int_{t}^{ \delta}dt'\,  [ c_3\cos m_a(t'-t) +(m_a^{-1}+c_2) \sin m_a(t'-t)] \sin  m_at'   \Big],\\
\frac{d\thb_{\rm in}(t)}{dt}  = &3m_aAH_N(0)  \Big[\int_{0}^{t}dt'\, [c_1 \sin m_a(t'-t) -c_2 \cos m_a(t'-t)]\sin m_at' \\
    +&\int_{t}^{ \delta}dt'\,  [ c_3\sin m_a(t'-t) -(m_a^{-1}+c_2) \cos m_a(t'-t)] \sin  m_at'   \Big]  +  3m_aAH_N(0) ( c_1 -c_3)   \sin m_at  
 .\label{eq:c1}
} 
At $t=0$,
 \dis{
\thb_{\rm in}(0) =&3m_aAH_N(0)   \int_{0}^{ \delta}dt'\,  [ c_3\cos m_at' +(m_a^{-1}+c_2) \sin m_at'] \sin  m_at'  \\
=&\frac34AH_N(0)    [ c_3(1-\cos 2m_a\delta)+(m_a^{-1}+c_2)(2-\sin 2 m_a\delta)  ]   \\
 }
where we choose  for one oscillation
\dis{
c_2=-m_a^{-1}, \delta=\frac{2\pi}{m_a},\label{eq:delta}
}
such that $\thb_{\rm in}$ grows from 0. Then,
 \dis{
\dot\thb_{\rm in}(0)& = 3m_aAH_N(0)  \int_{0}^{ \delta}dt'\,   c_3  \sin^2  m_at'   =\frac{3c_3  AH_N(0) }{2} \pi,
}
where $c_3$ is not fixed. $ \thb_{\rm in}$ grows from $t=0$ to $t=\delta$,
\dis{
 \thb_{\rm in}(\delta)& = 3m_aAH_N(0)  \int_{0}^{\delta}dt'\, [c_1 \cos m_a(t'-\delta) + c_2\sin m_a(t'-\delta)]\sin m_at' \\& = 3m_aAH_N(0)  \int_{0}^{\delta}dt'\, [c_1 \cos m_a\delta \cos m_at' -c_1 \sin m_a\delta \sin m_at' \\
 &\qquad -m_a^{-1} \cos m_a\delta \sin m_at'+m_a^{-1}\sin m_a\delta \cos m_at']\sin m_at' \\  
 & = 3m_aAH_N(0)  \int_{0}^{\delta}dt'\, [( m_a^{-1}\sin m_a\delta+c_1  \cos m_a\delta)\frac{\sin 2m_a t'}{2} -( m_a^{-1}\cos m_a\delta +c_1  \sin m_a\delta) \frac{1-\cos 2m_a  t'}{2} ] \\  
 & = 3m_aAH_N(0)    \left[( m_a^{-1}\sin m_a\delta+c_1  \cos m_a\delta)
  \frac{1-\cos 2m_a \delta}{4m_a}  -( m_a^{-1}\cos m_a\delta +c_1  \sin m_a\delta) (\frac{\delta}{2}-\frac{\sin 2m_a\delta}{4m_a})\right] \\  
 & = -\frac{3 \pi AH_N(0)}{m_a} .  \label{eq:atDelta}
} 
where we used (\ref{eq:delta}). So, the amplitude $A$ decreases to $A(1-\frac{3 \pi  H_N(0)}{m_a})$ where we neglected O($H_N^2$).
 
Until  $t_{\rm now}\approx 4\times 10^{17}\,$s,  corresponding to $T_{\rm now}\simeq 2.35\times 10^{-4}\,\eVV$,  the axion with $m_a=10^{-4}\,\eVV$  oscillated $N_{\rm now}=0.6\times 10^{29}$ times. For  $N_{\rm now}$ oscillations from $t_f$, we multiply the following $N_{\rm now}$ factors to obtain the current $\bar\theta$,  
\dis{
\frac{\thb_{\rm now}}{\thb_f}=\prod_{N=1}^{N_{\rm now}} \left(1-\frac{3 \pi  H_N(0)}{m_a}\sqrt{\frac{ 1}{N}} \right),
\label{eq:factNow}
}
which cannot be performed in simple ways. A different method will be used in Sec. \ref{sec:tbar}, after discussing the linearization regime before $t_f$.
 
\section{Waiting for linearization regime:  bottle neck period}\label{sec:AxCDMbottle} 
   
Consider the evolution of axion field $A=f_a \theta$ in the standard big Bang cosmology,  
\dis{
 \ddot{A}+3H\dot{A}+\frac{m_0^2\,f_a}{2}\sin\theta=0. \label{eq:Axsinfn}
}   
Once the linearization is realized, $\sin\theta\simeq \theta$  is used for its evolution.
  In the nonlinear regime, however, the linearization is no longer valid and a proper treatment  is needed \cite{Bae08}. Namely, we must include the dissipation term, which postpones effectively the beginning of oscillation and lowers the temperature of the  commencement time of oscillation. With the above sinusoidal form,  
let us introduce a conformal time $\tau(t)$ as a function of $t$ such that the first derivative term of $\theta$ with respect to $\tau$ in the differential equation (\ref{eq:diffeq}) is absent.  Then, we obtain
 \dis{
 \theta^{''}  +\frac{m^2}{\dot\tau^2}\sin\theta=0,\label{eq:diffeqtau}
}
where prime denote the derivative with respect to $\tau$ and $\ddot\tau/\dot\tau=-3H$, and 
 \dis{
m^2=\frac{m_0^2}{2}.\label{eq:massbelowone}
 }

Since the scale factor $a(t)$ has a power law with respect to $t$ in the radiation dominated (RD) and matter dominated (MD) universes, let us parametrize the Hubble parameter as $H=\dot{a}/a=n/t $. Then, 
 \dis{
 &\dot\tau=\exp\left[-3\int^t \,H(t')dt' \right]\nonumber \\[0.3em]
 &\quad=\left\{
\begin{array}{l} 
e^{-3Ht}~~\textrm{for constant~} H,\\[0.5em]
(t/t_0)^{-3n}~~\textrm{with } n=\frac12\textrm{ in RD and~}  n=\frac23\textrm{ in MD} .
\end{array}
\right.
 }
Thus,
 \dis{
\tau =\left\{
\begin{array}{l} 
-\frac{1}{3H}e^{-3Ht}~~\textrm{for constant~} H,\\[0.5em]
\frac{t_0}{1-3n}(t/t_0)^{-3n+1}~~\textrm{with } n=\frac12, \frac23\textrm{ respectively in the RD and MD universes}.  
\end{array}
\right.
 }
which gives  
 \dis{
 -2t_0\left(\frac{t}{t_0}\right)^{-1/2} \textrm{~and~}  - t_0\left(\frac{t}{t_0}\right)^{-1}, \nonumber 
 }
in the RD and MD Universes, respectively. At arount $T\lesssim 1\,\gev$, we will use $n=\frac12$  the RD  Universe value in Sects. \ref{subsec:Harmonic}  and \ref{subsec:constantmbar}.   Thus, $\theta$ evolves according to  the parametrization $H =1/2t$.  The $t$ range $(0,+\infty )$ is mapped into  the $\tau$ range  $(-\infty,0)$.

We use the following $t$ dependence of $\bar{m}$,
 \dis{
  \theta^{''}  + \bar{m}^{\,2}(\tau)\sin\theta=0,~~\textrm{with~}~
   \bar{m}^{\,2}(\tau)= 2^{21.32} m_0^2 \left(\frac{\tau}{t_0}\right)^{-22.32},
 \label{eq:maqphase}
 }
 where $'$ denotes derivative with respect to $\tau$. In this section, we introduce the bottle neck period analytically, starting with a constant mass.
  
In the following two subsections,  we discuss the evolution for a constant $\bar{m}$, firstly on the harmonic solution and second on the asymptotic analytic solution. Then,  In Sec. \ref{sec:tbar}, the time dependence of $m$ is used.

\subsubsection{To the linearization regime with constant $\bar{m}$ }\label{subsec:Harmonic}
  
Before considering the $\tau$ dependence of $\bar{m}$, let us consider it a constant $\bar{m}$ to acquaint with the bottle neck period.
When $\theta$ falls down to a sufficiently small value, we can approximate $\sin\theta\simeq\theta$, and Eq. (\ref{eq:maqphase}) is simplified to
\dis{
 \theta^{''}  +\bar{m}^{\,2}\theta=0.
 \label{eq:lineardiff}
  }
  In the limit $\tau\to-\infty$, $\theta$ approaches to a constant $\theta_0$. The linearization solution is to disregard this constant and we expand the linearization solution as descending functions of $\theta_k$
 \dis{
 \theta(\tau)=\sum_{k=0}^\infty\, \theta_k(\tau).
  }
Assuming $\bar{m} (\tau)$ is sufficiently small compared to $\tau^{-1}$, we obtain recursion relations,
\dis{
 &\theta_1''=-\bar{m}^{\,2}(\tau) \theta_0,\nonumber \\
 &\theta_2''=-\bar{m}^{\,2}(\tau) \theta_1(\tau),\nonumber \\
 &\quad\quad\cdots \nonumber\\
 & \theta_{k+1}''=-\bar{m}^{\,2}(\tau) \theta_k(\tau),\nonumber 
  }
 and the Neumann series solution is
\dis{
  \theta_{k+1}(\tau)= -\int_{-\infty}^{\tau}d\tau' \int_{-\infty}^{\tau'}d\tau''\, \bar{m}^{\,2} (\tau'')\theta_k(\tau''). 
  }
Since $\theta_0$ is taken as a constant, $\theta_k(\tau)$ can be expressed as a multiple integral,
\dis{
  \theta_{k}(\tau) =~ & \theta_0\int_{-\infty}^{\tau}d\tau_{2k} \int_{-\infty}^{\tau_{2k}}d\tau_{2k-1}\, [-\bar{m}^{\,2} (\tau_{2k-1})]  \\
  & \times \int_{-\infty}^{\tau_{2k-1}}d\tau_{2k-2} \int_{-\infty}^{\tau_{2k-2}}d\tau_{2k-3}\, [-\bar{m}^{\,2} (\tau_{2k-3})]   \\
  & \cdots  \\
  & \times \int_{-\infty}^{\tau_{3}}d\tau_{2} \int_{-\infty}^{\tau_{2}}d\tau_{1}\, [-\bar{m}^{\,2} (\tau_{1})] 
 \\
  =~ &(-1)^{k} \theta_0\left\{ \prod_{j=1}^k\int_{-\infty}^{\tau_{2j+1}}d\tau_{2j} \int_{-\infty}^{\tau_{2j}}d\tau_{2j-1}\, [\bar{m}^{\,2} (\tau_{2j-1})] \right\}_{\tau_{2k+1}=\tau}.
 }
For $n>\frac13$, with $\alpha=1/(3n-1)$, $\theta_k$ is given by

\begin{eqnarray}
\theta_k(\tau) &=&  \theta_0\,\frac{\Gamma(\frac{1}{2\alpha}+1)}{\Gamma(k+1)\Gamma(\frac{1}{2\alpha}+k+1)}\left[-\frac14 m^2 t_0^2 \left(\frac{|\tau|}{\alpha t_0}\right)^{-2\alpha} \right]^k,\\[0.7em]
\theta_k(t) &=& \theta_0\,\frac{\Gamma(\frac{1}{2\alpha}+1)}{\Gamma(k+1)\Gamma(\frac{1}{2\alpha}+k+1)}\left[-\frac14 m^2 t ^2   \right]^k.
\end{eqnarray}
By Mathematica, one can sum the series to find the hypergeometric function of the first kind,
\dis{
\theta(t)= ~_0F_1\left(\frac{1}{2\alpha}+1,-\frac14 m^2t^2\right)~\theta_0.\label{eq:hyperLinear}
 }
 
\subsubsection{Approximate solution in the (anharmonic) nonlinear regime}\label{subsec:constantmbar}

\begin{figure}[!t]
 \includegraphics[width=0.6\textwidth]{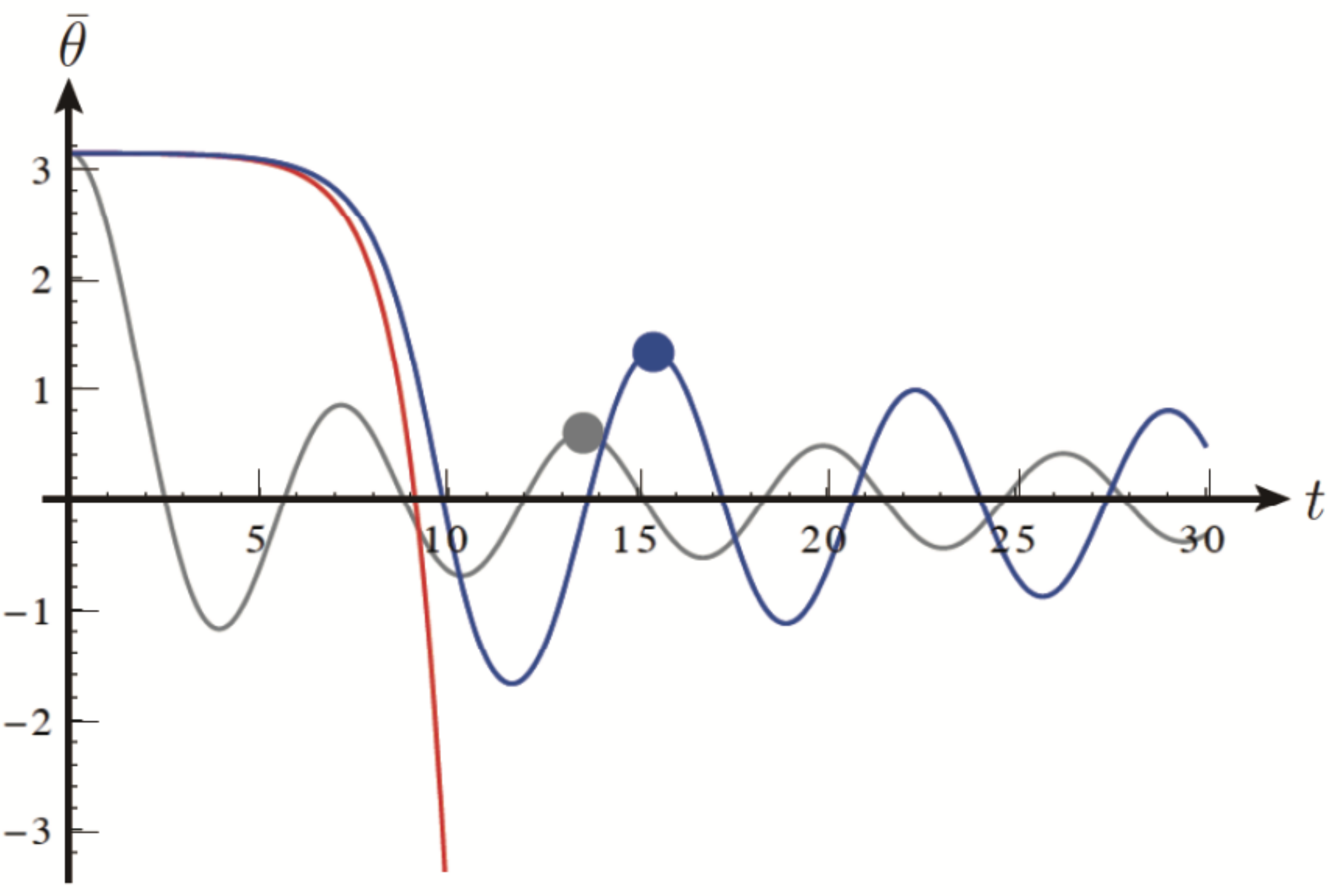}
\caption{ Equation (\ref{eq:anhAsymp}) is shown with the red color. It is compared with the numerical solution (in blue) of \cite{Huh16} and a solution for the harmonic potential (in gray), staring in the anharmonic regime $\theta_0=0.99\pi$ at $T=1\,\gev$. }\label{fig:AnharmOld}
\end{figure}

With the trick shown above, let us now study the region with a large value of $\theta_0$, \ie in a deep anharmonic regime,
\dis{
\theta_0\gg \sum_{k=1}^\infty\,\theta_k(\tau).
 }
Then, $\sin(\theta_0+\sum_{k=1} \,\theta_k(\tau))\simeq \sin\theta_0+\cos\theta_0\sum_k\theta_k$. Thus, the recursion relations are
\dis{
 &\theta_1''=-\bar{m}^{\,2}(\tau) \sin\theta_0,\nonumber \\
  &\theta_2''=-\bar{m}^{\,2}(\tau)\cos\theta_0\theta_1(\tau),\nonumber \\
  &\quad\quad \cdots \nonumber\\
 & \theta_{k+1}''=-\bar{m}^{\,2}(\tau)\cos\theta_0 \theta_k(\tau),\nonumber 
 }
leading to 
 \dis{
 \theta_k(\tau)  =(-1)^{k} \cos^{k-1}\theta_0\sin\theta_0 
 \cdot\left\{ \prod_{j=1}^k\int_{-\infty}^{\tau_{2j+1}}d\tau_{2j} \int_{-\infty}^{\tau_{2j}}d\tau_{2j-1}\, [\bar{m}^{\,2} (\tau_{2j-1})] \right\}_{\tau_{2k+1}=\tau}\,. 
 }
An asymptotic solution in the linear regime is expressible in terms of a hypergeometric function,
\begin{eqnarray}
\theta(t)_{\rm asymptotic}= \theta_0-\tan\theta_0\left[-1+ ~ _0F_1\left(\frac{5}{4},-\frac14 m^2 t^2\cos\theta_0\right) \right].\label{eq:anhAsymp}
\end{eqnarray}

Two solutions given in Eqs. (\ref{eq:hyperLinear}) and (\ref{eq:anhAsymp}) are shown in Fig. \ref{fig:AnharmOld}, in gray and red colors, together with the numerical solution of \cite{Bae08}.  We can express the solution in the entire region by two hypergeometric functions by joining two at a reasonable $\theta_k$ where the analytic solutions almost match the numerical solution.

\section{Numerical solution for the $\thb$ evolution}
\label{sec:tbar}
 
\begin{figure}[!t]
 \includegraphics[width=0.65\textwidth]{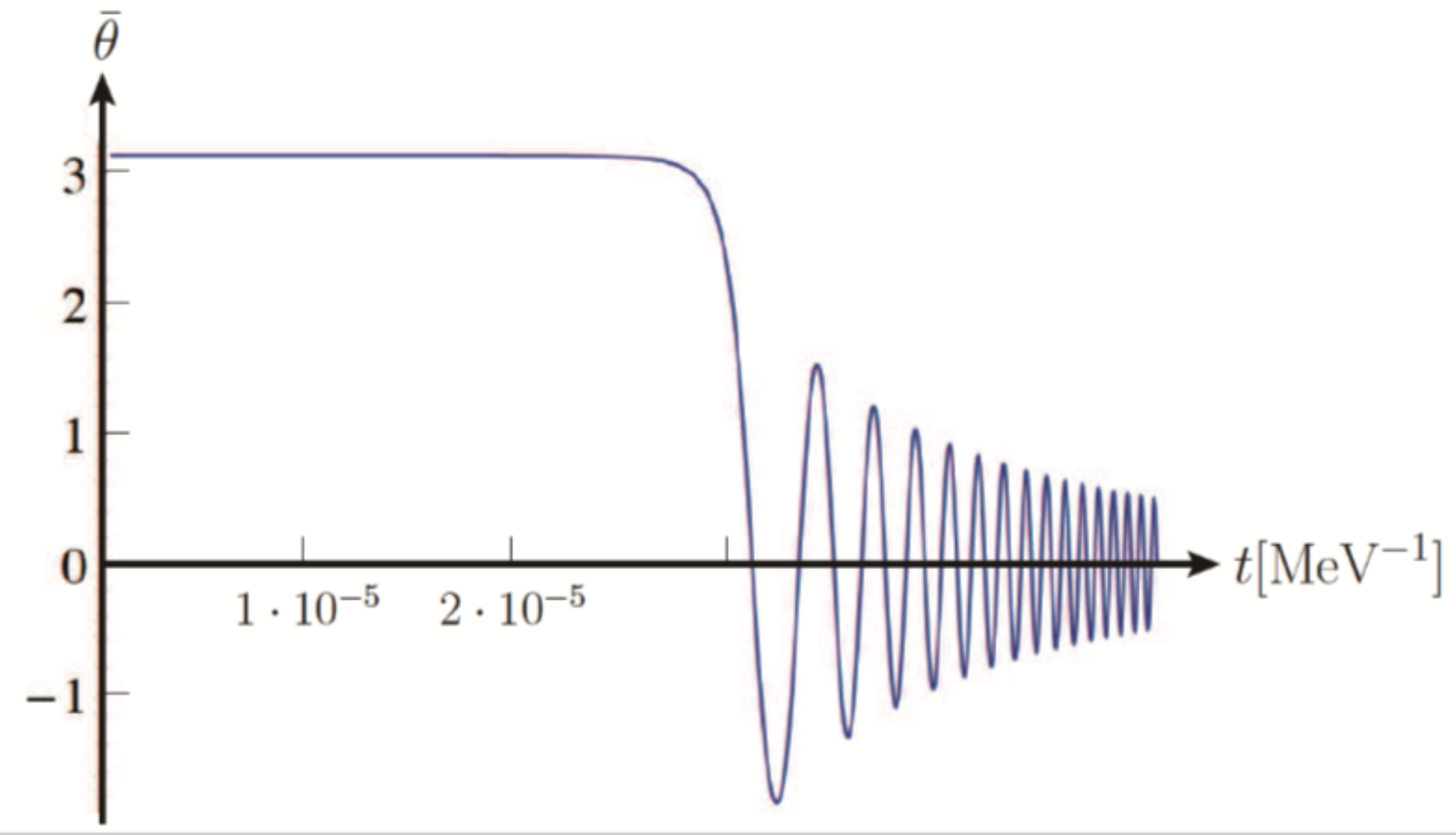}
\caption{$\bar\theta$ oscillation for the temperature dependent axion mass $m_a\simeq 10^{-4\,}\eVV$  at $T=0$ with $m_a(T)\propto T^{-8.2}$, starting in the anharmonic regime $\theta_0=0.99\pi$ at $T=1\,\gev$.  }\label{fig:Numerical}
\end{figure}

In fact, the bottle neck effect is more pronounced than that commented for a constant $\bar {m}$ in the previous subsection. As temperature is increased, the axion mass tends to zero very fast, which implies a much longer period for one oscillation. In other words, as temperature drops, the oscillation period becomes shorter and shorter. This behavior is shown in Fig. \ref{fig:Numerical} for $m_a=10^{-4\,}\eV$, starting with $\theta_1=0.99\,\pi$. This is because of the high power temperature dependence of the axion mass: $m_a\propto T^{-8.2}$.  Between the $\rho$ meson mass scale and $T_c$, we used the phenomenological power $\propto T^{-4.403}$ to match smoothly to the zero temperature value at $T_c$ \cite{KimKim18}. The numerical solution is shown in Fig. \ref{fig:Numerical}. We note that from the first maximum the anharmonic effects are negligible and we conclude that the bottleneck period has been finished after the first maximum.

In Ref. \cite{KimKim18}, the axion mass dependence of $\thb$ during the QCD phase transision has been given  as
\begin{equation}
r_{f/1}\equiv \frac{\thb_f}{\thb_1}\simeq 0.02\left(\frac{m_a}{10^{-4}\eVV}\right)^{-0.591\pm 0.008}.
\end{equation}

In Eq. (\ref{eq:factNow}), the number of multiplication  factors for cosmologically interesting axions was shown, which cannot be calculated  by direct computation.   For  $m_a\simeq  10^{-4\,}\eVV$, the  axion field oscillation is 152 GHz, having oscillated $\approx 0.65\times 10^{29}$ times until now. We need to find out a method to deal with this huge number of oscillations.


\begin{figure}[!t]
\includegraphics[width=14cm,height=7cm,
angle=0]{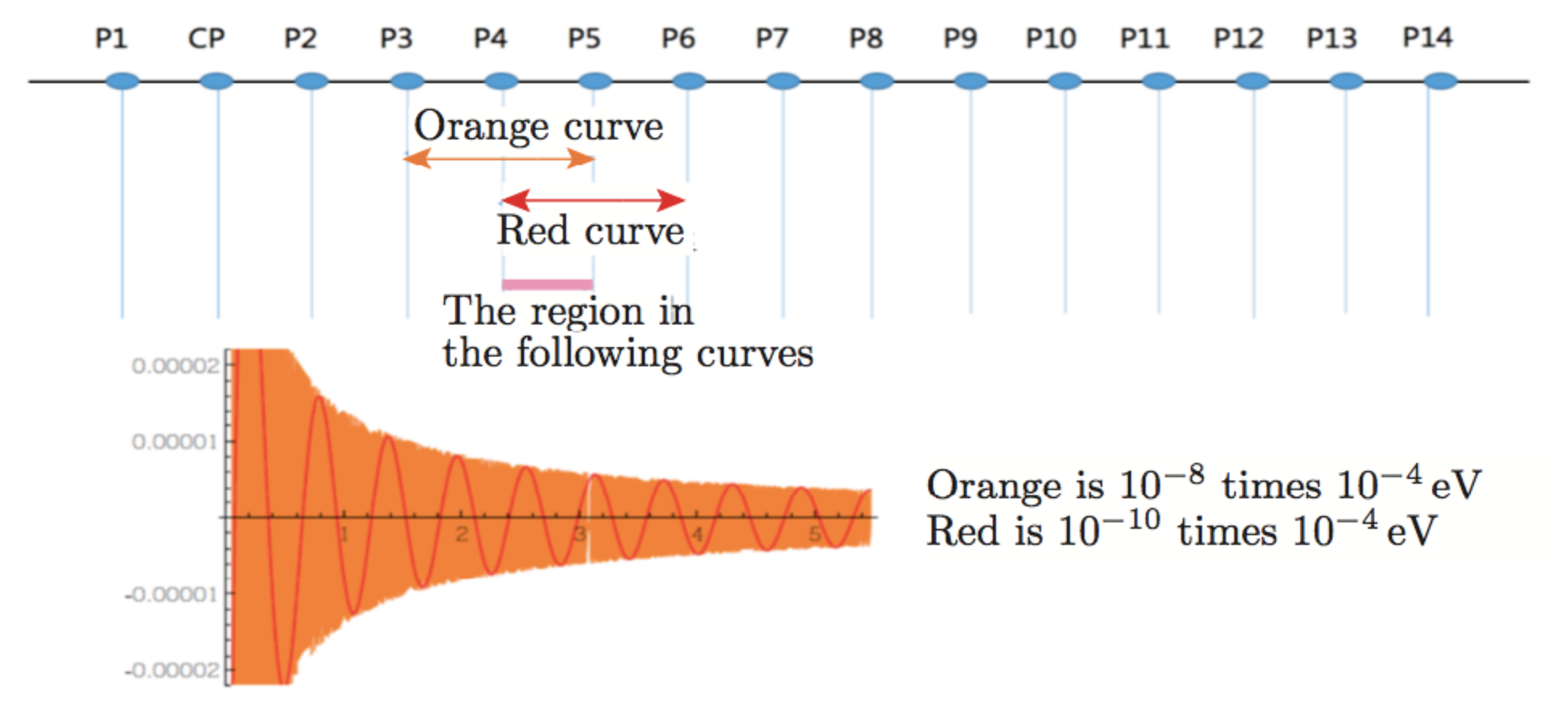}
 \caption{Comparison of $\thb$ evolutions for axion masses of $10^{-12}\eVV$ and $10^{-14}\eVV$. In each step between the points from $P2$ to $P14$ are differing by $10^2$. $CP$ is the point at the critical temperature. $P1$ is the point at $T_1$.   In the lower part, the orange curves are overlapping in our thickness of curves, and the unit of the horizontal axis is in seconds and the vertical value for $\thb$ is in the log scale. $\thb$ oscillation is calculated in the orange region with $m_a=10^{-12}\eVV$ and in the red  region with $m_a=10^{-14}\eVV.$  The oscillation figures   in the overlapping region shown as the lavender bar between $P4$ and $P5$ are used in our calculation. This process of overlapping is continued from the steps [$P1$ and $P2$] to [$P13$ and $P14$].}\label{fig:Compare}
\end{figure}

After $t_f$,  the evolution equation  (\ref{eq:Axsinfn}) can be approximated by the linear differential equation
\dis{
 \ddot{\thb}+3H\dot{\thb}+\frac{m_0^2}{2} \thb\simeq 0. \label{eq:AxLinear}
}   
Equation (\ref{eq:AxLinear}) has a symmetry
\dis{
&m_0\to 10^{-n}m_0\\
& \thb\to 10^{n}\thb.\label{eq:factor}
}
So, if we derease the axion mass by a factor $10^{-n}$, the needed number of oscillations will be decreased by a factor  $10^{-n}$. Then, the calculated $\thb$ must be  $10^{+n}$ times larger than the unit in the  previous oscillation. Let us choose $n=2$. Even though the linear equation (\ref{eq:AxLinear}) is useful to observe the factor increase of $\thb$ shown in Eq. (\ref{eq:factor}), we solve the exact equation numerically, rather than solving Eq. (\ref{eq:AxLinear}).
This method of decreasing the axion mass is confirmed in Fig.  \ref{fig:Compare} where two evolutions with two different axion masses differing by the factor $10^{2}$ are shown. In  Fig.  \ref{fig:Compare}, the reference time points are marked from $P1$ to $P14$.  $P1$ is the point at $T_1$, and  $CP$ is the point at the critical temperature.  The factor increase from $P2$ to $P14$ are by $10^2$.  From the point $P4$ to $P5$, the $\thb$ evolutions corresponding to axion masses  $10^{-12}\eVV$ and $10^{-14}\eVV$ are explicitly presented. In the lower part of Fig.   \ref{fig:Compare}, the  $\thb$ scale in the $y$-axis is logarthmic. 

\begin{figure}[!t]
 \includegraphics[width=0.65\textwidth]{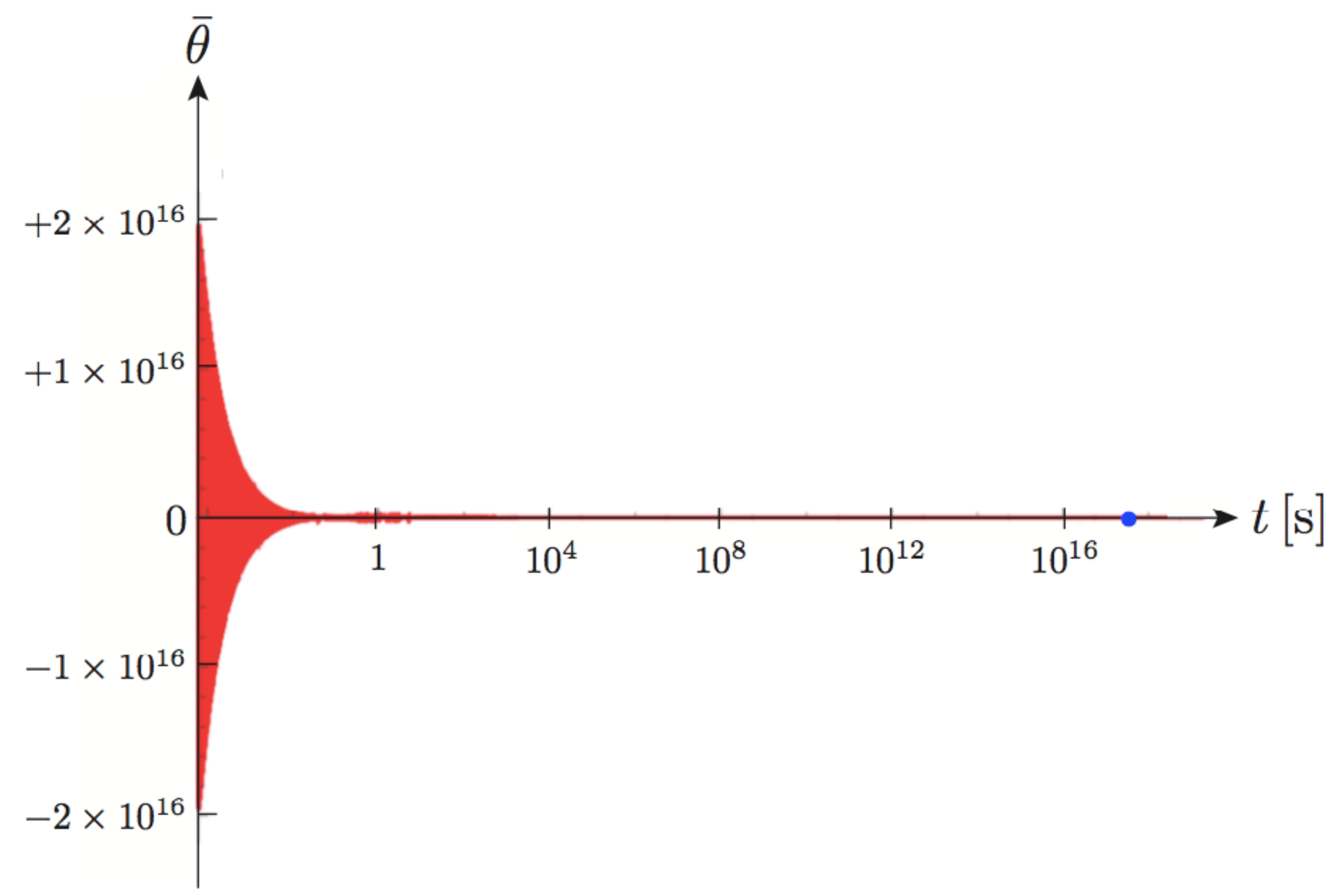}
\caption{$\bar\theta$ oscillation from $t_f$ until now.  The value at $t_f$ is $4.5\times 10^{16}$}\label{fig:RNowf}
\end{figure}

We repeat this process of decreasing axion mass using $n=2$ in each step and find out the increased $\thb$ at the present time $t_U\simeq 4\times 10^{17}$ s, which is shown in Fig.  \ref{fig:RNowf}. 
In this way,    for $\thb_1=1$ and  $m_a=10^{-4}\eVV$ we obtain,
\dis{
\frac{\thb_{\rm now}}{\thb_f}\simeq 3.07758 \times 10^{-17} .\label{eq:thbrationow}
}
Thus, we obtain $r_{\rm now/1}\simeq 6\times 10^{-19}$ for $m_a=10^{-4}\eVV$, and the current axion energy density
 as a function of $\thb_1$ near $m_a=10^{-4}\,\eVV$  is
\dis{
\frac{\rho_a}{[\eVV^4]}\simeq  5.68\cdot 10^{-6}\, \thb_1^2\left( \frac{m_a}{10^{-4}\,\eVV}\right)^{-3.182\pm 0.016}\simeq 2.1\cdot 10^{-6}\, \thb_1^2\left( \frac{f_a}{10^{11}\,\gev}\right)^{3.182\pm 0.016}
,  \label{eq:rhonow}
}
where we included the $\chi$ decrease factor 2 in the exponent.  Equation
(\ref{eq:rhonow})  can be compared to the current critical energy density $\rho_c\simeq 10^{-11}\,\eVV^4$. If 27\,\% of critical energy density is QCD axions, we need $\thb_1\simeq 10^{-3}$ and $3\times 10^{-5} $, respectively, for $f_a=10^{11}\,\gev$ and $10^{12}\,\gev$. These numbers prefer higher axion masses compared to the previous estimates \cite{KimRMP10}.
This study may alleviate the tension created by the axions created from the string-wall system \cite{Shellard10,Kawasaki12}.

\section{Conclusion}
\label{sec:Conclusion}
The evolution of misalignment angle $\thb$  after passing through the bottle neck period is calaculated. Knowledge of this evolution is crucial to determine the current energy density of cold axions. The onset of $\thb$ oscillation after the initial value $\thb_1$ at temperature $T_1$, satisfying $m_a(T_1)=3H(T_1)$, undergoes a bottleneck period which occurred during the QCD phase transition. In addition, the anharmonic coupling of order $a^4$ affects the $\thb$ evolution. The time $t_{\rm osc}$ that the anharmonic effect is negligible is taken as the onset of the first oscillation after the bottle neck period. From that time, the finishing time $t_f$ of the cross over phase transition was calculated before \cite{KimKim18}. We calculate the evolution of $\thb_{\rm now}$ from $t_f$ based on our observation of the symmetry (\ref{eq:factor}) with the ratio O($10^{-17}$) presented in Eq. (\ref{eq:thbrationow}). This enables us to estimate the axion energy density in terms of the initial misalignment angle $\thb_1$.

 \section*{Acknowledgments}
J.E.K. thanks Deog Ki Hong, Duchul Kim, and A. Ringwald for helpful communications. This work is supported in part by the National Research Foundation (NRF) grant funded by the Korean Government (MEST)
(NRF-2015R1D1A1A01058449). J.E.K. is supported also in part by IBS-R017-D1-2018-a00, and S. K.  supported in part also by NRF-2015aR1D1A1A09059301. 


\end{document}